\documentclass[12pt]{article}

\usepackage{amsmath,amsfonts,amssymb}
\usepackage{amsthm}
\usepackage{booktabs}
\usepackage{graphicx}
\graphicspath{{plots/}}
\usepackage{multirow}
\usepackage{algorithm,algpseudocode}
\usepackage[margin=1in]{geometry}
\usepackage[numbers,sort&compress]{natbib}
\usepackage{xcolor}
\definecolor{myblue}{RGB}{30 144 255}
\definecolor{mygreen}{RGB}{0 100 0}
\usepackage[colorlinks=true,citecolor=myblue,linkcolor=myblue,urlcolor=myblue]{hyperref}
\usepackage{xr}
\usepackage[doublespacing]{setspace}

\theoremstyle{plain}
\newtheorem{theorem}{Theorem}
\newtheorem{proposition}[theorem]{Proposition}
\theoremstyle{definition}

\newtheorem{definition}{Definition}
\newtheorem{assumption}{Assumption}
\theoremstyle{remark}
\newtheorem{remark}{Remark}

\newcommand{\bbeta}{\boldsymbol{\beta}}
\newcommand{\bzeta}{\boldsymbol{\zeta}}
\newcommand{\balpha}{\boldsymbol{\alpha}}
\newcommand{\bmu}{\boldsymbol{\mu}}

\newcommand{\btheta}{\boldsymbol{\theta}}

\newcommand{\bX}{\boldsymbol{X}}
\newcommand{\bx}{\boldsymbol{x}}

\newcommand{\by}{\boldsymbol{y}}

\newcommand{\bI}{\mathbf{I}}

\title{Bayesian Environment Invariant Regression}
\author{
Ruqian Zhang\thanks{Department of Statistics and Data Science, Fudan University, Shanghai, China.}
\and
Juan Shen\footnotemark[1]
\and
Yijiao Zhang\thanks{Department of Biostatistics, Epidemiology and Informatics, University of Pennsylvania, Philadelphia, PA, USA. Corresponding author: \href{mailto:yijiao.zhang@pennmedicine.upenn.edu}{yijiao.zhang@pennmedicine.upenn.edu}.}
}
\date{}

\begin{document}

\maketitle

\begin{abstract}
The availability of data from multiple heterogeneous environments has motivated methods that remain reliable under distributional shifts. When the joint distribution of response and predictors varies across environments, the response may still depend on a subset of predictors through an invariant mechanism. 
Existing methods typically assess candidate invariant sets through pooled stability criteria, treating environmental variation as nuisance.
In this paper, we propose a Bayesian framework that explicitly separates a shared response mechanism from environment-specific or response-dependent associations, exploiting heterogeneity as evidence for structure learning.
A competitive spike-and-slab prior is designed to force each predictor to compete between invariant and non-invariant spurious effects.
Under a tractable working model, we establish invariant model selection consistency and posterior contraction for invariant coefficients.
We further study the presence of irrelevant predictors, characterize posterior concentration on an equivalent invariant class, and introduce a post-selection refinement that consistently recovers the minimal invariant model. Simulations and a real application illustrate the robustness and finite-sample efficiency of the proposed method.
\end{abstract}


\section{Introduction}

Data collected from multiple environments are common in scientific studies, where combining related yet heterogeneous data is often essential for efficient inference and prediction.
Environmental perturbations can alter the joint distribution of response and predictors, violating the classical assumption that training and test samples are identically distributed.
In the presence of such heterogeneity, models trained within a single environment may exploit spurious associations that fail to generalize under distribution shifts, leading to poor out-of-distribution performance.

A well-known illustration is the cow-camel image classification example in \citet{arjovsky2020irm}.
In Internet images, both background color and body shape can be predictive of the animal label.
A classifier trained in that environment may use background color as a shortcut and achieve strong in-distribution accuracy.
Yet when it is deployed on farm images with nearly homogeneous backgrounds, predictive performance can deteriorate sharply even though body shape remains informative. 
This example highlights a statistical challenge: some predictors are linked to a stable causal mechanism governing the response, whereas others are predictive solely due to environment-specific associations with the response.

Robust learning from multiple environments has therefore received increasing attention in statistics and machine learning \citep{Meinshausen2015,Guo2024maximin}.
In many applications, scientific knowledge suggests that, despite distributional heterogeneity, the response depends on a subset of predictors through a mechanism that remains stable across environments. 
This observation motivates the pursuit of invariant learning, which aims to identify stable structure in heterogeneous environments and to uncover causal response mechanisms.

\subsection{Related work}

A growing literature has studied invariant learning from different perspectives.
The seminal work of \citet{ICP2016} adopts a causal framework and identifies invariant predictors as direct causes of the response in structural causal models.
They propose a hypothesis testing procedure for detecting predictor subsets whose regression residual distributions remain invariant across environments. 
This approach has been extended to nonlinear settings \citep{HeinzeDeml2018} and to sequentially ordered data \citep{Pfister2019}.
While conceptually appealing, testing-based approaches can be conservative for model selection and may lose power in finite samples \citep{Fan2024Environment}.

An alternative line of work learns invariance through regularized optimization.
Anchor regression \citep{anchor2021} augments pooled empirical risk with a penalty on projected residuals after regressing out candidate predictors. \citet{Fan2024Environment} study invariant conditional expectation in multi-environment linear least squares by introducing penalties on correlations between residuals and predictors; related nonlinear extensions are discussed in \citet{Gu2025}. \citet{Shen2026causal} further consider invariant gradient learning, trading off in-distribution prediction and causality-oriented invariance.
These methods generally treat heterogeneity as a constraint to be controlled via tuning parameters, so their practical performance and statistical guarantees depend on appropriate penalty calibration.

Within the Bayesian paradigm, \citet{wu2025bayesian} propose a Bayesian invariant prediction framework that expresses invariance through structured probabilistic models. They then perform posterior inference over candidate invariant sets based on pooled conditional likelihood. 
Despite these advances, many existing approaches continue to evaluate candidate invariant sets through pooled performance or pooled stability criteria, treating environment-specific variation mainly as nuisance to be neutralized rather than as information to be modeled. 
This raises a fundamental question: can environmental heterogeneity itself be exploited as a statistical resource for identifying invariant structures?

We argue that environmental heterogeneity can be leveraged through modeling the joint distribution, rather than being eliminated. 
From this perspective, our key task is to disentangle invariant effects in the shared response mechanism from spurious effects, which arise through environment-specific or response-dependent associations. 
Consequently, instead of relying solely on the predictive aspect of invariance, we aim to identify predictors that participate in the shared mechanism, which is the object of causal interest.

\subsection{Our approach: Bayesian structural competition}

We propose \textbf{B}ayesian \textbf{E}nvironment \textbf{I}nvariant \textbf{R}egression (\texttt{BEIR}), a Bayesian framework that simultaneously models the invariant response mechanism and the spurious associations within a joint probabilistic formulation. The invariant structure is represented through latent indicators. 
Unlike the standard Bayesian variable selection framework, where a binary indicator only determines whether a predictor is active \citep{george1993variable, Ishwaran2005ss}, our indicator specifies the structural role of a predictor: it is either assigned a common coefficient in the invariant response model, or assigned environment-specific coefficients in the response-dependent association model. We implement this idea through a competitive spike-and-slab prior that induces block-wise competition between these two mutually exclusive explanations for each predictor. Posterior inference therefore balances cross-environment stability against environment-specific flexibility, turning heterogeneity into posterior evidence for structural learning.

Under this formulation, our results yield the following theoretical insights. First, heterogeneity can facilitate identification of the structural invariant set, including settings where homogeneous environments lead to non-identifiability. Second, under a simplified working model designed for scalable posterior inference, both invariant set selection consistency and true invariant coefficient estimation consistency are guaranteed. 
Third, when environment-specific and response-dependent associations are ignored, the pooling-based Bayesian inference can be systematically biased towards including predictors outside the shared response mechanism. These results formalize environmental heterogeneity as a valuable resource rather than a nuisance.

We further study the practically important case involving irrelevant predictors, which have no association with the response in all environments.
In the presence of such predictors, the invariant set is no longer uniquely defined, but forms an equivalent invariant set class consisting of supersets of the true structural invariant set, which may include irrelevant predictors.
We establish posterior concentration on this equivalent class and characterize a phase transition in posterior preference within the class driven by the prior hyperparameters.
Moreover, we propose a refinement procedure (\texttt{BEIR+}), which combines invariant screening (\texttt{BEIR}) with an additional feature selection step to consistently recover the minimal invariant set, which only contains all truly influential predictors in the invariant response mechanism.

\subsection{Organization}
 
Section~\ref{sec:prob_setup} formalizes invariant learning via probabilistic factorization and presents a Gaussian linear instantiation, together with a scalable working model.
Section~\ref{sec:BEIR} proposes \texttt{BEIR}, including the competitive spike-and-slab prior and the posterior inference via a Gibbs sampler.
Section~\ref{sec:thm} establishes theoretical guarantees of \texttt{BEIR}.
Section~\ref{sec:irre_predictors} extends the methodology to a general case with irrelevant predictors and introduces the refinement procedure \texttt{BEIR+}.
Section~\ref{sec:numerical} reports simulation studies and the Sachs' protein signaling application.
Concluding remarks are given in Section~\ref{sec:discussion}.

\section{Problem Setup}
\label{sec:prob_setup}

\subsection{Structural invariant learning from a probabilistic factorization}
\label{sec:invariant_setting}

Let $\mathcal E$ be the environment set with its size $E:=|{\mathcal E}|$.
Let ${\mathcal D}^{(e)}=\{(y_i^{(e)},\bx_i^{(e)})\}_{i=1}^{n^{(e)}}$ be i.i.d. samples from environment $e$, with response $y_i^{(e)}\in\mathbb{R}$ and predictors $\bx_i^{(e)}\in\mathbb{R}^p$, and ${\mathcal D}=\{{\mathcal D}^{(e)}\}_{e\in{\mathcal E}}$.
We allow for heterogeneity across environments, where the joint distribution $p^{(e)}(y^{(e)},\bx^{(e)})$ may vary with $e$, capturing possible distribution shifts induced by interventions or unobserved contextual changes.
A broad notion of predictive invariance characterizes stable association at the conditional distribution level and assumes that there exists an index set $I\subseteq[p]$ such that, for any $e,e'\in{\mathcal E}$,
\begin{equation*}
\label{eq:invariant_set}
    p^{(e)}(y^{(e)} \mid \bx^{(e)}_I) = p^{(e^\prime)}(y^{(e^\prime)} \mid \bx^{(e^\prime)}_I),
\end{equation*}
where $\bx^{(e)}_I$ denotes the subvector of $\bx^{(e)}$ corresponding to $I$.
However, predictive invariance may include variables whose stable association with the response arises through response-dependent pathways, rather than the shared response mechanism of interest.
For example, downstream readouts in biological systems or post-outcome variables in economic studies may appear predictively invariant without actually being part of the underlying scientific mechanism.

Motivated by this association distinction, we adopt a more conservative and mechanism-oriented notion of invariance through a probabilistic factorization of the joint distribution. Specifically, for any invariant set $I$ and any environment $e\in{\mathcal E}$,
\begin{equation}
\label{eq:factorization}
p^{(e)}(y^{(e)},\bx^{(e)}_{I},\bx^{(e)}_{I^c})
=
p^{(e)}(\bx^{(e)}_{I^c}\mid y^{(e)},\bx^{(e)}_I)
g(y^{(e)}\mid \bx^{(e)}_{I})
p^{(e)}(\bx^{(e)}_I),
\end{equation}
where the conditional distribution $g(y^{(e)} \mid \bx_I^{(e)})$ is shared across ${\mathcal E}$ and other predictors are modeled via $p^{(e)}(\bx_{I^c}^{(e)} \mid y^{(e)}, \bx_I^{(e)})$, which may vary with $e$ or depend on $y^{(e)}$.
Such a structure inspires an invariant spirit focusing on the response mechanism, which strengthens purely predictive invariance and aligns closely with the causal interpretation emphasized by \citet{ICP2016}.

\begin{definition}[Structural invariant set]
\label{def:invariant_set}
    
    An index set $I \subseteq [p]$ is called a structural invariant set if there exists some $g(\cdot)$ such that, for all $ e \in \mathcal E$,
    \begin{equation*}
        p^{(e)}(y^{(e)} \mid \bx_I^{(e)}) = g(y^{(e)} \mid \bx_I^{(e)}),
    \end{equation*}
    with $\{x^{(e)}_j\}_{j\in I}$ independent of the noise in the data-generating process of $y^{(e)}$.
\end{definition}

Compared with predictive invariance, a structural invariant set excludes variables whose stable association with the response is induced through response-dependent behavior, and therefore targets a more mechanism-based concept of invariance. 
Predictors in $I$ are termed structural invariant predictors, whereas predictors in $I^c$ are referred to as spurious predictors. 
The spurious set $I^c$ includes both environment-heterogeneous predictors and response-dependent predictors. 
Accordingly, our goal is to identify predictors that contribute to the invariant response mechanism while screening out predictors whose associations are not structurally stable.

The factorization in Eq.\eqref{eq:factorization} accommodates a broad class of response mechanisms. In this paper, we adopt a Gaussian linear representation for all components to simplify presentation and enhance interpretability.
In this section, we focus on the setting where all predictors are associated with the response, which will be relaxed in Section~\ref{sec:irre_predictors}.
Given the invariant set $I$, for all $j\in I^c$, we model the predictors $x^{(e)}_j$ to explicitly depend on $y$ by
\begin{equation}
\label{eq:X_model_full}
x_{j}^{(e)} = \mu_{j}^{(e)} + \alpha_{j}^{(e)}\,y^{(e)} + (\bx^{(e)}_I)^T \bzeta^{(e)}_j + \xi_{j}^{(e)},
\quad \xi_{j}^{(e)}\sim N(0,(\sigma_{x,j}^{(e)})^2),
\end{equation}
where $\mu_j^{(e)}$ is the intercept, $\bzeta^{(e)}_j$ denotes the dependence between spurious and invariant predictors given the response, and $\alpha_j^{(e)}$, referred to as the spurious coefficient, captures the association between $x_j^{(e)}$ and $y^{(e)}$ with at least one $\alpha_j^{(e)}\neq 0$ for $e\in{\mathcal E}$. 
The invariant response mechanism is modeled by
\begin{equation}
\label{eq:y_model}
y^{(e)} = (\bx_I^{(e)})^T \bbeta_I + \varepsilon^{(e)},
\quad \varepsilon^{(e)}\sim N(0,\sigma_y^2),
\end{equation}
where $\bbeta_I$, referred to as invariant coefficient, is shared across environments and $\beta_j\neq 0$ for $j\in I$.
The non-zero assumptions in Eq.\eqref{eq:X_model_full}-\eqref{eq:y_model} represent a multi-environment version of causal minimality to guarantee identifiability \citep{Peters2013identify,Park2020identify}.
To complete the joint distribution, for invariant predictors $x^{(e)}_j$ for $j\in I$, we specify
\begin{equation}
\label{eq:X_I_model}
x_{j}^{(e)} = \mu_{j}^{(e)} + \xi_{j}^{(e)},
\quad \xi_{j}^{(e)}\sim N(0,(\sigma_{x,j}^{(e)})^2).
\end{equation}
Under the data-generating process of $(y^{(e)},\bx^{(e)})$ for $e \in {\mathcal E}$ defined in Eq.\eqref{eq:X_model_full}-\eqref{eq:X_I_model}, we denote the true structural invariant set as $I_0$.

While $\bx^{(e)}_{I^c}$ are allowed to depend on $\bx^{(e)}_I$ in the true joint distribution, fitting the full model would substantially increase model complexity. Since our objective is to capture the invariant response mechanism rather than the spurious predictors, we consider a tractable working specification that retains the response-dependent component of the association structure.
We keep Eq.\eqref{eq:y_model} and \eqref{eq:X_I_model} unchanged and consider a simplified version of Eq.\eqref{eq:X_model_full} as the working spurious model to be fitted, where for $j\in I^c$,
\begin{equation}
\label{eq:X_model}
x_{j}^{(e)} = \mu_{j}^{(e)} + \alpha_{j}^{(e)}\,y^{(e)} + \xi_{j}^{(e)},
\quad \xi_{j}^{(e)}\sim N(0,(\sigma_{x,j}^{(e)})^2).
\end{equation}
This working model reduces the complexity for modeling spurious predictors and enables scalable implementation.
We denote parameters under $I$ in the working model by
$\btheta_{(I)} = (\bbeta_I, \sigma^2_y, \{\balpha^{(e)}_{I^c}, \bmu^{(e)}, \Sigma^{(e)}_{x} \}^E_{e=1})$,
with $\bmu^{(e)}=(\mu^{(e)}_1,\ldots,\mu^{(e)}_p)^T$ and
$\Sigma^{(e)}_x=\operatorname{diag}((\sigma^{(e)}_{x,1})^2, \ldots, (\sigma^{(e)}_{x,p})^2)$.
Let $|I|$ denote the size of $I$.

\begin{remark}[Working conditional independence]
\label{remark:condindep}
The working model \eqref{eq:X_model} assumes that, conditional on $y^{(e)}$, each $x^{(e)}_{j}$ for $j\in I^c$ is independent of $\bx^{(e)}_{I}$.
This simplification yields a scalable computation scheme, whose robustness and computational efficiency are examined in simulation studies.
As will be shown in Theorem \ref{thm:isc} in Section \ref{sec:thm}, the proposed inference procedure based on the working likelihood can still consistently recover the true invariant set $I_0$ defined through Eq.\eqref{eq:X_model_full}-\eqref{eq:X_I_model} under certain conditions.

\end{remark}

\subsection{Identifiability of structural invariance}
\label{sec:identify}

In this subsection, we study the identifiability of the true structural invariant set.
Specifically, under the true data-generating mechanism defined in Eq.\eqref{eq:X_model_full}-\eqref{eq:X_I_model}, we examine whether the structural invariant set is identifiable by the joint distribution, which justifies it as a meaningful scientific target.
In the presence of multiple environments, we have the following results.

\begin{proposition}[Identifiability under the true model]
\label{prop:true_iden}

Under the true model class defined in Eq.\eqref{eq:X_model_full}-\eqref{eq:X_I_model}, an invariant set $I$ is identifiable, if for any $\tilde{I} \neq I$: 

\begin{enumerate}
    \item[(a)] If $\tilde{I}\subset I$, assume $\exists e, e^\prime \in {\mathcal E}$ such that $\sum_{j\in I\setminus \tilde{I}}\beta_j \mu^{(e)}_j \ne \sum_{j\in I\setminus \tilde{I}}\beta_j \mu^{(e^\prime)}_j$ or $\sum_{j\in I\setminus \tilde{I}} \beta_j^2 (\sigma^{(e)}_{x,j})^2 \ne \sum_{j\in I\setminus \tilde{I}} \beta_j^2 (\sigma^{(e^\prime)}_{x,j})^2$.

    \item[(b)] If $I\subset \tilde{I}$, assume $\exists e, e^\prime \in {\mathcal E}$ such that $\sum_{j\in \tilde{I} \setminus I}\frac{(\alpha^{(e)}_j)^2}{(\sigma^{(e)}_{x,j})^2} \ne \sum_{j\in \tilde{I} \setminus I}\frac{(\alpha^{(e^\prime)}_j)^2}{(\sigma^{(e^\prime)}_{x,j})^2}$.

    \item[(c)] If neither $\tilde{I}\subset I$ nor $I\subset \tilde{I}$, assume $\exists e, e^\prime \in {\mathcal E}$ such that $(\boldsymbol{M}^{(e)})^{-1}\boldsymbol{c}^{(e)} \ne (\boldsymbol{M}^{(e^\prime)})^{-1}\boldsymbol{c}^{(e^\prime)}$ with $\boldsymbol{M}^{(e)}$ and $\boldsymbol{c}^{(e)}$ defined in Eq.(S1.3) and (S1.4), respectively.
\end{enumerate}

\end{proposition}

Proposition~\ref{prop:true_iden} establishes identifiability of the scientific target under the true joint model in Eq.\eqref{eq:X_model_full}-\eqref{eq:X_I_model} and highlights the role of cross-environment heterogeneity in identifying the true structural invariant set $I_0$. 
Intuitively, assumption (a) ensures that omitting true invariant predictors breaks invariance across environments; 
assumption (b) ensures that including additional spurious predictors introduces environment heterogeneous associations; assumption (c) rules out the case where omitted invariant signals can be exactly compensated for by spurious predictors. 
These assumptions are sufficient with detailed arguments in Section~S1.1.

In Section~S1.2 of the supplementary materials, we consider the simplified joint distribution in Eq.\eqref{eq:y_model}-\eqref{eq:X_model} as a special case to more transparently illustrate the necessity of cross-environment heterogeneity. 
Proposition~S1.1 demonstrates that introducing heterogeneous environments can resolve the intrinsic non-identifiability under homogeneous environments in certain scenarios, which is empirically illustrated in Section~\ref{sec:demo}.

\section{Bayesian Environment Invariant Regression}
\label{sec:BEIR}

We propose \texttt{BEIR}, a Bayesian method that fits the working model and recovers the true invariant structure by jointly exploiting shared and spurious associations.

\subsection{Competitive spike-and-slab prior}
\label{sec:Bayes}

The joint distribution of $(y^{(e)},\bx^{(e)}_I,\bx^{(e)}_{I^c})$ encodes invariance through a structural factorization by $I$,
which motivates us to treat $I$ as a structural indicator and perform invariant learning by letting each predictor compete between an invariant role and a spurious role.
We introduce binary indicators $I_j\in\{0,1\}$ for $j\in[p]$, where $I_j=1$ indicates that predictor $x_j$ is modeled by Eq.\eqref{eq:y_model} and belongs to the invariant set $I=\{j: I_j=1\}$, and $I_j=0$ indicates that it is assigned to the response-dependent modeling defined in Eq.\eqref{eq:X_model}.

To capture this structural information, we introduce a competitive spike-and-slab prior based on $I_j$.
This prior has two complementary components that jointly model invariant and spurious effects.
Conditional on $I$, for each $j\in[p]$,
\begin{equation}
\label{eq:beta_prior}
	\beta_j\mid I_j\stackrel{ind}{\sim} I_j N(0,{\sigma^{2}_y}\tau^2)+(1-I_j)\delta_0,
    \quad I_j\stackrel{iid}{\sim} \operatorname{Bern}(\gamma),
\end{equation}
where $\delta_0$ denotes the point mass at $0$, $\tau$ controls the slab variance for invariant effects, and $\gamma$ is the prior inclusion probability.
The point-mass spike prior ensures that the nonzero components of $\bbeta$ coincide with $I$ in the posterior distribution.

The indicator $I_j$ simultaneously governs the response-dependent models \eqref{eq:X_model}.
For each $e\in\mathcal E$ and $j\in[p]$, the prior on spurious coefficient $\alpha^{(e)}_j$ is specified as
\begin{equation}
\label{eq:alpha_prior}
	\alpha^{(e)}_j\mid I_j \stackrel{ind}{\sim} I_j\delta_0+(1-I_j)N(0,(\sigma^{(e)}_{x,j})^2\eta^2),
\end{equation} 
where $\eta$ controls the slab variance for spurious effects.
Thus, $I_j=1$ forces $\alpha_j^{(e)}\equiv 0$ for all $e$, whereas $I_j=0$ forces $\beta_j\equiv 0$.
The joint prior then factorizes as
\begin{equation*}
\pi(I,\bbeta,\{\balpha^{(e)}\}_{e=1}^E)
=
\prod_{j=1}^p
\left[
\pi(I_j)\,\pi(\beta_j\mid I_j)\,
\prod_{e=1}^E \pi(\alpha_j^{(e)}\mid I_j)
\right].
\end{equation*}

Unlike conventional variable selection, where an indicator controls whether a coefficient is active, the priors in Eq.\eqref{eq:beta_prior}-\eqref{eq:alpha_prior} allow $I_j$ to select between two mutually exclusive parameterizations: a shared invariant coefficient versus $E$ environment-specific coefficients.
In this way, the posterior of each $I_j$ no longer evaluates whether a predictor is useful, but instead induces a block-wise competition for each $x_j$, between an invariant explanation and a spurious explanation on its association with the response.
Rather than solely focusing on the invariant likelihood $g(y\mid \bx_I)$, the posterior directly incorporates cross-environment variation into inference and utilizes heterogeneous information to distinguish invariant response mechanism.

\begin{remark}[Structural competition]
\label{rem:competitive}

In the competitive prior, the invariant mode activates one shared parameter $\beta_j$, while the spurious mode activates $E$ environment-specific parameters $\alpha_j^{(e)}$.
The two modes differ in their complexity.
Consider a standardized setting. The log-posterior contributions under two modes are $\log\gamma-\frac{1}{2}\log\tau^2$ under $I_j=1$ and $\log(1-\gamma)-\frac{E}{2}\log\eta^2$ under $I_j=0$ up to constants.
The preference of assigning $x_j$ to the invariant role rather than the spurious role is 
\begin{equation*}
    \Delta
:=
\log\frac{\gamma}{1-\gamma}
-\frac{1}{2}\log\tau^2
+\frac{E}{2}\log\eta^2.
\end{equation*}
A positive value of $\Delta$ indicates a higher complexity cost for the spurious mode and a stronger prior preference towards the invariant role.
This complexity difference reveals how the slab hyperparameters, $\eta$ and $\tau$, affect the structural allocation.

\end{remark}

\subsection{Joint posterior and Gibbs sampler}

For nuisance parameters, we adopt conjugate priors, where the priors for the noise variances are specified as inverse gamma distributions ${\sigma^2_y}\sim \operatorname{IG}(a_0,b_0)$ and $(\sigma^{(e)}_{x,j})^2\sim \operatorname{IG}(c_0,d_0)$, respectively, and the priors for the intercepts are set as Gaussian distributions $\mu^{(e)}_j\sim N(0,(\sigma^{(e)}_{x,j})^2\sigma^2_{\mu})$ for $j\in[p]$ and $e\in \mathcal{E}$, with hyperparameters $a_0$, $b_0$, $c_0$, $d_0$, and $\sigma^2_\mu$. The full joint prior is denoted as $\pi(I, \btheta_{(I)} )$.
Under the working model \eqref{eq:y_model}-\eqref{eq:X_model}, the joint likelihood factorizes as
\begin{equation*}
	\begin{aligned}
		L_n({\mathcal D}\mid I, \btheta_{(I)})&=\prod^E_{e=1}\prod^{n^{(e)}}_{i=1} p^{(e)}(\bx^{(e)}_{i, I^c}\mid y^{(e)}_i, \bx^{(e)}_{i,I}) g( y^{(e)}_i \mid  \bx^{(e)}_{i,I}) p^{(e)}(\bx^{(e)}_{i,I}),
	\end{aligned}
\end{equation*}
and the joint posterior distribution follows from the Bayes' theorem as
\begin{equation*}
	\begin{aligned}
		&\pi(I, \btheta_{(I)}\mid {\mathcal D})\propto L_n({\mathcal D}\mid I, \btheta_{(I)})\pi(I, \btheta_{(I)} ).
	\end{aligned}
\end{equation*}

Posterior inference is carried out via a Gibbs sampler.
Invariant selection is performed based on the posterior probability of each $I_j$.
Notably, the simplification under our working model avoids sampling $\bzeta^{(e)}_{j}$ for $j\in [p]$ and $e \in {\mathcal E}$.
Detailed updates for all parameters are deferred to Section~4 of the supplementary materials.

\section{Theoretical Guarantees}
\label{sec:thm}

\subsection{Posterior behavior of \texttt{BEIR}}

We now establish theoretical guarantees for \texttt{BEIR}.
Our main results characterize the behavior of the working posterior in Eq.\eqref{eq:y_model}-\eqref{eq:X_model} while the true data-generating mechanism is allowed to be more general.
In particular, we study whether the working posterior still concentrates on the true structural invariant set $I_0$ and the true invariant coefficients $\bbeta_{0,I_0}$ defined by Eq.\eqref{eq:X_model_full}-\eqref{eq:X_I_model}, where $\bx^{(e)}_{j}$ depends on $\bx^{(e)}_{I_0}$ through $\bzeta^{(e)}_{0,j}$ in the true spurious mechanism for $j\in I_0^c$ and $e\in{\mathcal E}$.
To simplify exposition, we assume known noise variances and omit intercepts by centering, which does not affect the conclusions.
Throughout this section, we work under the identifiability conditions in Section~\ref{sec:identify}, so that the true invariant set $I_0$ is unique.

Let $(\by^{(e)},\bX^{(e)})$ denote the response vector and design matrix in environment $e$.
We stack response vectors and design matrices across $\mathcal E$ and denote them as
$\tilde{\by}=((\by^{(1)})^T,\ldots,(\by^{(E)})^T)^T$ and
$\tilde{\bX}=((\bX^{(1)})^T,\ldots,(\bX^{(E)})^T)^T$, respectively,
with total sample size $N=\sum_{e=1}^E n^{(e)}$.
For any candidate invariant model $I\subseteq[p]$, let $\tilde{\bX}_I$ denote the stacked design restricted to coordinates in $I$.

We first provide regularity assumptions for posterior concentration under the competitive spike-and-slab formulation.

\begin{assumption}
\label{assump:1}
	For each $e \in \mathcal{E}$,
	(i) $n^{(e)}/N \ge C_N$ for some constant $C_N\in (0,1)$,
	(ii) $\|{\bX^{(e)}}\bbeta_{0} \|^2_2 \le C_1 n^{(e)}\sigma_y^2$ for some constant $C_1 >0 $.
\end{assumption}

Assumption~\ref{assump:1}(i) prevents any environment from being negligible, which is common in multi-source analysis \citep{Duan2023,Zhang2025Concert}.
Assumption~\ref{assump:1}(ii) gives an upper bound on the environment-wise total true signal strength \citep{yang2016on}.

\begin{assumption}
\label{assump:2}
	For any candidate model $I$ and some constants $0<\lambda_{l}<\lambda_{u}<\infty$, (i) the stacked Gram matrix satisfies $\lambda_{l}\le \lambda_{\min}(\frac{\tilde{\bX}_I^T\tilde{\bX}_I}{N}) \le \lambda_{\max}(\frac{\tilde{\bX}_I^T\tilde{\bX}_I}{N}) \le \lambda_{u}$,
    and (ii) for each $e\in \mathcal E$, $\lambda_{l}\le \lambda_{\min}(\frac{(\bX^{(e)}_I)^T\bX^{(e)}_I}{n^{(e)}}) \le \lambda_{\max}(\frac{(\bX^{(e)}_I)^T\bX^{(e)}_I}{n^{(e)}}) \le \lambda_{u}$.
\end{assumption}

Assumption~\ref{assump:2} is a multi-environment version of restricted eigenvalue conditions and ensures identifiability for candidate models \citep{Narisetty2019ss}.

\begin{assumption}
\label{assump:3}
	For some constants $C_\tau, C_{\eta} \ge 1$ and $C_\gamma >0$,
	(i) $\tau^2N \asymp N^{C_\tau}$, (ii) $\eta^2 N \asymp N^{C_\eta}$, (iii) $\gamma/(1-\gamma) =C_\gamma$.
\end{assumption}

Assumption~\ref{assump:3} specifies the scaling of the spike-and-slab hyperparameters and matches standard regimes in Bayesian variable selection \citep{bondell2012}.

\begin{assumption}
\label{assump:4}
	For some constant $C_\beta, C_{\alpha}, C_u, C_{\zeta}>0$,
	(i) $\min_{j\in I_0} |\beta_{0,j}|^2 \ge C_{\beta}\sigma^2_y \log N/N$,
	(ii) $\min_{j\in I_0^c} \sum^E_{e=1}|\alpha^{(e)}_{0,j}|^2/(\sigma^{(e)}_{x,j})^2\ge C_{\alpha} \log N/N$,
    (iii) $\|\bbeta_{0,I_0}\|_{\infty} \le C_u$,
    and (iv) $|\alpha^{(e)}_{0,j}|^2  \sigma^2_y \ge C_{\zeta }\|\bzeta^{(e)}_{0,j}\|^2_2$ for all $j\in I^c_0$ and $e\in{\mathcal E}$.
\end{assumption}

Assumption~\ref{assump:4}(i)-(ii) impose beta-min type conditions for separating invariant and spurious mechanisms under posterior competition.
The spurious minimum signal strength condition is required at the cross-environment level, which is weaker than the single-environment analogue \citep{buhlmann2011statistics}.
Assumption~\ref{assump:4}(iii) excludes diverging invariant coefficients \citep{Raskutti2011}. Assumption~\ref{assump:4}(iv) ensures that the additional dependence of spurious predictors on invariant predictors is dominated by the response-dependent signal.

\begin{theorem}[Invariant model selection consistency]
\label{thm:isc}
	Under Assumptions~\ref{assump:1}-\ref{assump:4}, the working posterior probability of the true structural invariant model $I_0$ satisfies
	\begin{equation*}
		\Pi(I_0\mid {\mathcal D}) \ge \frac{1}{1+ c_1N^{-c_2}},
	\end{equation*}
	with probability at least  $1-c_3 e^{-c_4 N}$ for some constants $c_1, c_2, c_3, c_4 >0 $.
\end{theorem}

Theorem~\ref{thm:isc} shows that, even when the true spurious likelihood is misspecified, the working posterior continues to concentrate on the true structural invariant set, which justifies the use of the simplified working model \eqref{eq:X_model}.
Since the working model only simplifies the modeling of spurious variables, while preserving the target invariant mechanism of $y$, consistent identification of $I_0$ further ensures consistent estimation of the true invariant coefficients through the working posterior, as established in the following theorem.
Let $\bbeta_0\in \mathbb{R}^{p}$ denote the full coefficient vector, with subvector $\bbeta_{0,I_0}\in \mathbb{R}^{|I_0|}$ on $I_0$ and zeros on its complement $I_0^c$.

\begin{theorem}[Estimation of invariant coefficients]
	\label{thm:est}
	Under Assumptions~\ref{assump:1}-\ref{assump:4}, the posterior induced by the working model satisfies
	\begin{equation*}
		\Pi\left( \|\bbeta-\bbeta_{0}\|_2\ge \sqrt{\frac{M\sigma^2_y|I_0|\log N}{N}}\mid {\mathcal D} \right) \le c_5N^{-c_6}
	\end{equation*}
	with probability at least $1-c_7|I_0|N^{-c_8} $, for some constants $M,c_5,c_6,c_7,c_8>0$.
\end{theorem}

Theorem~\ref{thm:est} shows that the working posterior continues to allocate most of its mass around the true $\bbeta_{0}$ despite the simplified working specification for spurious predictors in Eq.\eqref{eq:X_model}. 
The posterior contraction rate depends on the total sample size $N$, reflecting that invariant effects are learned by integrating evidence across environments under the shared response mechanism.

\subsection{Model selection bias from pooling}

We contrast \texttt{BEIR} with a pooling-based analysis that ignores the structural factorization in Eq.\eqref{eq:factorization} and instead fits a single regression of $y$ on all predictors.
Specifically, consider the misspecified joint distribution
\begin{equation*}
	p^{(e)}(y^{(e)}, \bx^{(e)})=g(y^{(e)}\mid \bx^{(e)})p^{(e)}( \bx^{(e)}),
\end{equation*}
in which, under the linear regression setup, $g(y^{(e)}\mid \bx^{(e)})$ is $y^{(e)}=(\bx^{(e)})^T\bbeta+\varepsilon^{(e)}$.
Using the traditional spike-and-slab prior for $\bbeta$, for each $j\in[p]$,
\begin{equation*}
	\beta_j\mid I_j\stackrel{ind}{\sim} I_j N(0,{\sigma^{2}_y}\tau^2)+(1-I_j)\delta_0,
    \quad I_j\stackrel{iid}{\sim} \operatorname{Bern}(\gamma).
\end{equation*}
Under the true joint distribution in Eq.\eqref{eq:X_model_full}-\eqref{eq:X_I_model}, pooling fails to distinguish the invariant mechanism from the environment-specific mechanisms. 
Let ${\mathcal I}^{(e)}$ denote the sample indices from environment $e$, and $h_{ii}$ denote the $i$th diagonal element of $\tilde{\bX}_{I_0}(\tilde{\bX}_{I_0}^T\tilde{\bX}_{I_0})^{-1}\tilde{\bX}_{I_0}^T$.
The following result shows that, under a sufficient condition preventing cross-environment cancellation of spurious effects, model misspecification can lead to a systematic model selection bias towards including spurious predictors.

\begin{assumption}
\label{assump:mis_alpha}
For some constants $C_l, C_m, C_Z >0$ and each $j\in I^c_0$,
(i) $|\sum_{e=1}^E\alpha^{(e)}_{0,j} w^{(e)}|\ge C_l \sqrt{\log N/N}$ with $w^{(e)}=\sum_{i \in {\mathcal I}^{(e)} } (1-h_{ii})/N$,
(ii) $\max_{e\in{\mathcal E}}|\alpha^{(e)}_{0,j}|\le C_m$,
and (iii) $\sum^E_{e=1}\|\bzeta^{(e)}_{0,j}\|^2_2 \le C_Z$.
\end{assumption}

Assumption~\ref{assump:mis_alpha}(i) requires that the weighted average of spurious coefficients does not vanish, preventing their cancellation after pooling.
Assumption~\ref{assump:mis_alpha}(ii)-(iii) bound the spurious coefficients and dependence of $\bx^{(e)}_{I^c_0}$ on $\bx^{(e)}_{I_0}$, respectively.

\begin{proposition}
\label{prop:mis_model}

    Under Assumptions~\ref{assump:1}-\ref{assump:3} and \ref{assump:mis_alpha}, suppose $I^c_0\ne \emptyset$. The pooling posterior $\check{\Pi}(\cdot\mid \mathcal{D})$ satisfies
\begin{equation*}
	\frac{\check{\Pi}(I_0\mid {\mathcal D})}{\check{\Pi}([p]\mid {\mathcal D})} \stackrel{\mathrm{P}}{\longrightarrow} 0 \quad\text{as }N\to \infty.
\end{equation*}
\end{proposition}

Proposition~\ref{prop:mis_model} implies that, when the invariant structure is ignored, the pooled posterior places vanishing mass on the true invariant set relative to sets that include spurious predictors.
Whenever a spurious predictor exists, the pooling posterior tends to include it.
This finding underscores the necessity of explicitly modeling the invariant response mechanism together with the environment-specific structure.

\section{Refinement for Irrelevant Predictors}
\label{sec:irre_predictors}

Our earlier development assumes that each predictor belongs either to the invariant set or the spurious set.
In realistic applications, the observed predictors may also include features that are uninformative for response and unaffected by environmental perturbations.
To address the presence of irrelevant predictors, this section introduces a refinement procedure to recover the minimal invariant structure.

\subsection{Concentration on an equivalent invariant class}

Assume there exists a subset of indices $Z \subset [p]$ such that for each $j\in Z$,
\begin{equation*}
	\beta_{0,j}=0 \quad\text{and}\quad \alpha^{(e)}_{0,j}=0,\quad\forall e\in{\mathcal E}.
\end{equation*}
Predictors in $Z$ neither influence the response through the invariant mechanism nor are influenced by the response in the spurious model.
We define $Z$ as the \textit{irrelevant set}. Under this extension, the joint distribution in $e$ is augmented as
\begin{equation}
\label{eq:iref}
	p^{(e)}(y^{(e)},\bx^{(e)}_{I},\bx^{(e)}_{I^c},\bx^{(e)}_Z)=p^{(e)}(\bx^{(e)}_{I^c}\mid y^{(e)},\bx^{(e)}_I) g(y^{(e)}\mid \bx^{(e)}_{I}) p^{(e)}(\bx^{(e)}_I)p^{(e)}(\bx^{(e)}_Z).
\end{equation}
Since predictors in $Z$ are independent of both the invariant and spurious mechanisms, they can be viewed as structurally invariant across environments.
Consequently, the definition of invariant set in Definition~\ref{def:invariant_set} does not distinguish truly influential invariant predictors from irrelevant predictors.

Although \texttt{BEIR} remains applicable in the presence of $Z$, there are two concerns.
First, from the competitive prior, the indicator $I$ is designed to distinguish nonzero invariant and spurious effects, which is not directly defined for zero effects in $Z$.
Second, from the theoretical results, consistently identifying the invariant set requires the minimum signal strength on either $\beta_{0,j}$ or $\{\alpha^{(e)}_{0,j}\}_{e=1}^E$.
However, for irrelevant predictors in $Z$, both signals vanish.
These concerns raise the natural question of
how the posterior behaves in the presence of irrelevant predictors?

To construct an appropriate target set that incorporates irrelevant predictors in invariant model selection, we define the equivalent invariant class as
\begin{equation*}
	{\mathcal M}_{eq}=\{I: I_0\subseteq I\subseteq (I_0\cup Z) \}.
\end{equation*}
Each model in ${\mathcal M}_{eq}$ retains all truly influential invariant predictors and excludes any spurious predictors, while possibly including irrelevant predictors.
The following theorem shows that \texttt{BEIR} remains valid with its posterior concentrating on ${\mathcal M}_{eq}$.

\begin{theorem}[Equivalent invariant class selection consistency]
\label{thm:extended_isc}
	Under Assumptions~\ref{assump:1}-\ref{assump:4}, the working posterior probability of the equivalent invariant class ${\mathcal M}_{eq}$ satisfies
	\begin{equation*}
		\Pi({\mathcal M}_{eq}\mid {\mathcal D}) \ge \frac{1}{1+ c^\prime_1N^{-c^\prime_2}},
	\end{equation*}
	with probability at least $1-c^\prime_3 e^{-c^\prime_4 N}$ for some constants $c^\prime_1, c^\prime_2, c^\prime_3, c^\prime_4 >0 $.
\end{theorem}

Theorem~\ref{thm:extended_isc} implies that \texttt{BEIR} continues to separate the invariant mechanism from the spurious association with high probability. Any model selected from the posterior contains all predictors in $I_0$ and excludes spurious predictors in $I_0^c$.
The remaining ambiguity is confined to deciding whether the minimal invariant model $I_0$ is favored by the posterior within this class.
The subsequent result characterizes a phase transition governed by the relative scaling of the slab variances in Assumption~\ref{assump:3}.

\begin{proposition}[Selection within ${\mathcal M}_{eq}$]
	\label{prop:extended_I0}
	(i) If $C_\tau/E > C_\eta$, the posterior favors more parsimonious models. For any $I^\prime_0$ satisfying $I_0 \subseteq I^\prime_0 \subset (I_0\cup Z)$ and any $j\in Z\cap (I^\prime_0)^c$, 
	\begin{equation*}
		\frac{\Pi(I^\prime_0\cup \{j\}\mid {\mathcal D})}{\Pi(I^\prime_0\mid {\mathcal D})} \stackrel{\mathrm{P}}{\longrightarrow} 0 \quad\text{as } N\to \infty.
	\end{equation*}
	
	(ii) If $C_\eta > C_\tau/E +2$, the posterior favors larger models. For any $I^\prime_0$ satisfying $I_0 \subset I^\prime_0 \subseteq (I_0\cup Z)$ and any $j\in Z\cap I^\prime_0$,
	\begin{equation*}
		\frac{\Pi(I^\prime_0\setminus \{j\}\mid {\mathcal D})}{\Pi(I^\prime_0\mid {\mathcal D})} \stackrel{\mathrm{P}}{\longrightarrow} 0 \quad\text{as }N\to \infty.
	\end{equation*}
\end{proposition}

Proposition~\ref{prop:extended_I0} shows that within the equivalent class, posterior model preference can switch between parsimonious and expansive regimes.
This behavior aligns with the structural competition discussed in Remark~\ref{rem:competitive}, where adding an irrelevant predictor $j\in Z$ in $I$ activates an invariant slab coefficient, while the competing spurious mode would activate $E$ environment-specific coefficients.
As a result, the relative scaling of $(\tau,\eta)$ and the number of environments $E$ jointly determine whether the posterior tends to absorb irrelevant predictors into the invariant model.

\subsection{Post-selection refinement}

The asymptotic regimes in Proposition~\ref{prop:extended_I0} provide insights into hyperparameter effects, but finite-sample behavior may not exhibit a clear separation.
Motivated by Theorem~\ref{thm:extended_isc}, we propose a post-selection refinement strategy.

Since \texttt{BEIR} asymptotically isolates $\mathcal M_{eq}$, where any $I^\prime_0$ contains $I_0$ and excludes spurious predictors, the remaining task is to remove irrelevant predictors from $I^\prime_0$, which is a feature selection problem.
We therefore combine \texttt{BEIR} with a feature selection method ${\mathcal V}$, which can be any valid procedure, such as the lasso \citep{lasso1996} or spike-and-slab inference \citep{Ray2022vb}.
After running \texttt{BEIR}, we obtain a candidate model $\hat{I}$ from posterior inclusion probabilities.
We then apply method $\mathcal V$ on $\hat{I}$ to remove irrelevant predictors in $Z$.
This yields a family of hybrid procedures, denoted by \texttt{BEIR+}, which combines invariant learning with an additional sparsity refinement step. We summarize \texttt{BEIR+} in Algorithm~\ref{alg:beirplus}.

\begin{algorithm}[htbp]
\caption{Post-Selection Refinement: \texttt{BEIR+}}
\label{alg:beirplus}
\begin{algorithmic}[1]

\Statex \hspace{-\algorithmicindent}\textbf{Input:} Dataset $\mathcal D=\{(\by^{(e)},\bX^{(e)})\}_{e=1}^{E}$;
Spike-and-slab hyperparameters $(\tau,\eta,\gamma)$;
Posterior inclusion probability threshold $\kappa\in(0,1)$;
Feature selection method $\mathcal V$.
\Statex \hspace{-\algorithmicindent}\textbf{Output:} Refined invariant set $\hat I_{0}$.

\Statex\hspace{-\algorithmicindent}\textbf{Step 1: Invariant screening via \texttt{BEIR}}
\State Run \texttt{BEIR} via the Gibbs sampler on $\mathcal D$ and obtain the estimated posterior inclusion probabilities
$\hat\Pi(I_j=1\mid\mathcal D)$ for $j=1,\ldots,p$.
\State Define candidate set as $\hat I=\{j\in[p] :\hat\Pi(I_j=1\mid\mathcal D)\ge \kappa\}$.

\Statex\hspace{-\algorithmicindent}\textbf{Step 2: Post-selection feature selection}
\State Restrict the data to predictors in $\hat I$: $\mathcal D_{\hat I}=\{(\by^{(e)},\bX^{(e)}_{\hat I})\}_{e=1}^{E}$.
\State Apply $\mathcal V$ on $\mathcal D_{\hat I}$ to obtain the refined invariant set
$\hat I_{0}\subseteq \hat I$.

\Statex\hspace{-\algorithmicindent} \textbf{Return} $\hat I_{0}$.
\end{algorithmic}
\end{algorithm}

\begin{remark}[Guarantee of \texttt{BEIR+}]
\label{remark:beirplus}
By Theorem~\ref{thm:extended_isc}, $\hat I$ produced by \texttt{BEIR} belongs to ${\mathcal M}_{eq}$ with high probability.
Therefore, any method ${\mathcal V}$ that is model selection consistent on the reduced problem $\mathcal D_{\hat I}$ yields consistent recovery of the true invariant set $I_0$.
\end{remark}

\section{Numerical Studies}
\label{sec:numerical}

\subsection{Simulation}

In this section, we conduct simulation studies to evaluate the finite-sample performance of \texttt{BEIR} and \texttt{BEIR+}.
Unless otherwise specified, all results are averaged over 100 replications.

\subsubsection{Illustration of the role of heterogeneous environments}
\label{sec:demo}

We begin with a toy experiment to illustrate that adding an extra environment with potential heterogeneity is distinct from merely increasing the sample size.
Data are generated from the working distribution defined in Eq.\eqref{eq:y_model}-\eqref{eq:X_model} with $p=10$ and $n\in\{25,50,100\}$ observations per environment.
The true invariant set $I_0$ has size $s\in\{1,2,3\}$, and the true invariant coefficients are set as $\beta_{0,j}=0.5\epsilon_j$ for $j\in I_0$, where $\epsilon_j$ are independent Rademacher variables.
All remaining predictors are spurious and act through environment-specific $\{\alpha_{0,j}^{(e)}\}_{j\in I_0^c}$ without introducing $\{\bzeta^{(e)}_{0,j}\}_{j\in I_0^c}$.

We construct five environments with different levels of heterogeneity.
In the baseline environment \textbf{E1}, for each $j\in I_0^c$, we draw $\alpha_{0,j}^{(1)}\sim \mathrm{Unif}(1/2,1)\epsilon_j$. Environment \textbf{E2} is \emph{homogeneous} with E1 by setting $\balpha_{0,I_0^c}^{(2)}=\balpha_{0,I_0^c}^{(1)}$. Environment \textbf{E3} is \emph{strongly heterogeneous} by flipping all spurious signs, that is $\balpha_{0,I_0^c}^{(3)}=-\balpha_{0,I_0^c}^{(1)}$. Environments \textbf{E4} and \textbf{E5} represent \emph{moderate heterogeneity} via perturbations around E1, where
$\balpha_{0,I_0^c}^{(e)}\sim N(\balpha_{0,I_0^c}^{(1)},0.25^2\bI)$ for $e\in\{4,5\}$.
Intercepts $\mu^{(e)}_j$ are set as $0$ with $\sigma^{(e)}_{x,j}=\sigma_y=1$ for all $j\in [p]$ and $e\in {\mathcal E}$.
We enlarge the training set by sequentially incorporating environments from E1 to E5.

We compare \texttt{BEIR} to a pooled Bayesian spike-and-slab baseline (\texttt{Pool}; \citealp{Ray2022vb}), which ignores the invariant structure and fits a regression model on the aggregated data. We set hyperparameters $\tau=\eta=1$ and $\gamma =0.1$ throughout our simulation studies.
A sensitivity analysis on $(\tau,\eta)$ and the role of homogeneous environment splitting is reported in Section~5.1.

\begin{figure}[ht]
	\centering
	\includegraphics[width=\linewidth]{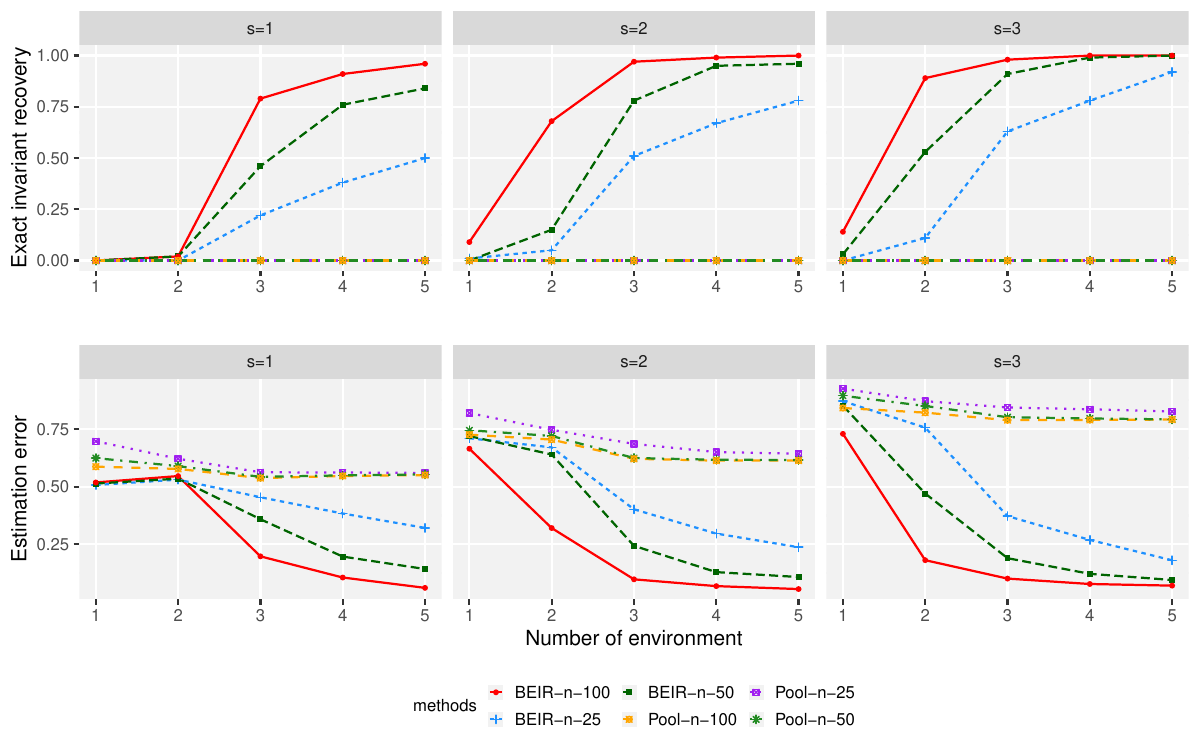}
	\caption{Toy simulation under different heterogeneous levels. Top: frequency of exact invariant set recovery. Bottom: $\ell_2$ estimation error of invariant coefficients.}
    \label{fig:set1_nonzero}
\end{figure}

Figure~\ref{fig:set1_nonzero} reports the exact recovery frequency of $I_0$ and the $\ell_2$ estimation error $\|\hat{\bbeta}-\bbeta_0\|_2$.
We observe three patterns.
First, \texttt{BEIR} improves as the number of environments increases, while \texttt{Pool} benefits little.
Note that the gain for \texttt{BEIR} is not solely explained by increased sample size.
The inclusion of the highly heterogeneous E3 yields a markedly greater improvement than adding an identical E2, even when the total sample size is smaller, for example, $E=3$ for $n=25$ (line {\color{myblue}\texttt{BEIR-n-25}}) versus $E=2$ for $n=50$ (line {\color{mygreen} \texttt{BEIR-n-50}}).
This highlights that heterogeneity in spurious mechanisms provides valuable information beyond what is obtained from homogeneous replication.

Second, when $s=1$, if only identical environments (E1 and E2) are observed, exact recovery remains poor regardless of $n$, reflecting the ambiguity discussed in Proposition~S1.1.
Once environment E3 is introduced, \texttt{BEIR} recovers $I_0$ with a higher frequency.
When $s\in\{2,3\}$, recovery becomes feasible even under identical environments as $n$ grows, consistent with the identifiability conditions for $|I_0|\ge2$ under Eq.\eqref{eq:y_model}-\eqref{eq:X_model}. This demonstrates the effectiveness of \texttt{BEIR} in screening out stable yet response-dependent predictors.

Finally, we observe possible improvement when adding the identical environment E2, part of which we attribute to the competitive prior.
Increasing the number of environments enlarges the spurious complexity penalty, encouraging the posterior to favor the invariant representation across environments.

\subsubsection{Performance without irrelevant predictors}
\label{sec:compare_no_zero}

We compare \texttt{BEIR} against pooling-based and invariant learning methods in settings without irrelevant predictors.
We set $p=10$, $n=50$, and $s=2$ with $\beta_{0,j}=0.5\epsilon_j$ for $j\in I_0$.
Spurious effects in the baseline environment E1 are $\alpha_{0,j}^{(1)}\sim \mathrm{Unif}(1/2,1)\epsilon_j$ for $j\in I_0^c$. For other environments $e\ge 2$, we perturb $\balpha^{(e)}_{0,I_0^c} = \balpha^{(1)}_{0,I_0^c} + r \boldsymbol{d}^{(e)}$,
where $r\in\{0.5,1,1.5,2\}$ controls the heterogeneity strength and $\boldsymbol{d}^{(e)}$ specifies the heterogeneity direction.
The number of environments is $E\in\{2,\ldots,10\}$.
If $E=2$, $\boldsymbol{d}^{(2)}$ is sampled uniformly from the unit sphere.
If $E>2$, we construct approximately orthonormal directions, which are centered around $\balpha^{(1)}_{0,I^c_0}$ with $\sum_{e=2}^E \boldsymbol{d}^{(e)}=0$ to ensure monotonically increasing heterogeneity.
For $j \in I_0^c$ and $e \in \mathcal{E}$, we sample $\bzeta_{0,j}^{(e)} \sim N(\mathbf{0}, 0.25^2 \bI)$ to introduce moderate dependence of $\bX^{(e)}_{I^c_0}$ on $\bX^{(e)}_{I_0}$. A more challenging case is provided in Section~S5.5 of the supplementary materials.
Each intercept $\mu^{(e)}_j$ is randomly sampled from $N(0,0.5^2)$ and the noise standard deviations are $\sigma^{(e)}_{x,j}=\sigma_y=1$.

We compare \texttt{BEIR} against pooled lasso (\texttt{Pooled\_lasso}; \citealp{lasso1996}) and spike-and-slab variational Bayes (\texttt{Pooled\_svb}; \citealp{Ray2022vb}), as well as invariant learning methods \texttt{EILLS} \citep{Fan2024Environment}, \texttt{ICP} \citep{ICP2016}, and \texttt{BIP} \citep{wu2025bayesian}.
All competing methods use default implementation configurations and recommended hyperparameter settings.

\begin{figure}[ht]
    \centering
    \includegraphics[width=\linewidth]{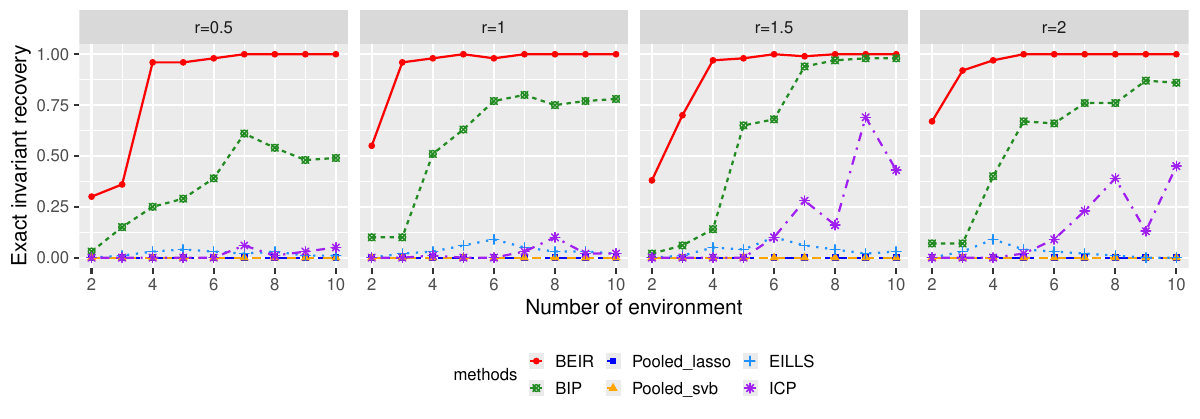}
    \caption{Frequency of exact invariant set recovery under different heterogeneity strengths $r$ and number of environments $E$ when the per-environment sample size is $n=50$.}
    \label{fig:set2_nonzero_acc}
\end{figure}

Figure~\ref{fig:set2_nonzero_acc} reports the exact recovery frequencies with estimation errors deferred to Section~S5.2.
Performance of invariant learning methods improves with both the number of environments and heterogeneity strength, reflecting accumulation of cross-environment evidence for stability.
Across all settings, \texttt{BEIR} achieves the highest recovery rate, with \texttt{BIP} ranking second, highlighting the advantage of explicitly encoding invariance through a structured probabilistic formulation.
These results also support the effectiveness of the working model despite its simplification.

\begin{figure}[ht]
    \centering
    \includegraphics[width=\linewidth]{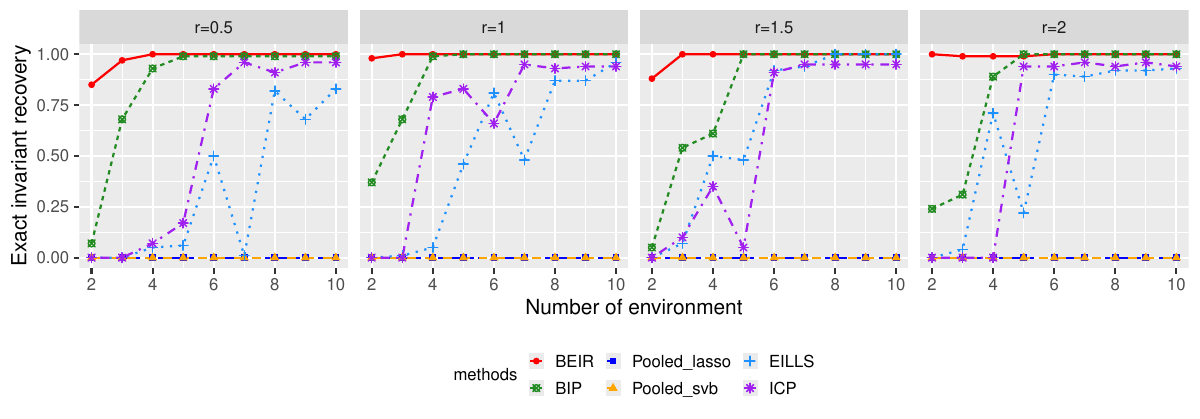}
    \caption{Frequency of exact invariant set recovery under different heterogeneity strengths $r$ and number of environments $E$ when the per-environment sample size is $n=200$.}
    \label{fig:set2_nonzero_acc_n200}
\end{figure}

We further increase the per-environment sample size to $n=200$ and provide the results in Figure~\ref{fig:set2_nonzero_acc_n200}.
All invariant learning methods improve substantially as $n$ increases, suggesting that sufficient sample sizes can alleviate the challenge for invariant learning. However, pooled methods show limited improvement, indicating that simply adding samples cannot compensate for structural misspecification.
Among invariant methods, \texttt{BEIR} continues to perform overall the best.
Notably, \texttt{BEIR} achieves comparable performance with fewer environments and smaller samples, demonstrating its ability to efficiently use cross-environment information.

\subsubsection{Performance with irrelevant predictors and refinement}
\label{sec:simu_zero}

In this subsection, we assess performance in the presence of irrelevant predictors and evaluate the refinement procedure \texttt{BEIR+}, with further investigation on the phase transition behavior of \texttt{BEIR} in Section~S5.3 of the supplementary materials.
We adopt the setting in Section~\ref{sec:compare_no_zero} with $r=2$ and $n=50$, and augment the predictor set by adding $d_Z=4$ Gaussian irrelevant predictors, which are independent of other components as in Eq.\eqref{eq:iref}, with $\bbeta_{0,Z}=\boldsymbol{0}$ and $\balpha^{(e)}_{0,Z}=\boldsymbol{0}$ for all $e$.

\begin{figure}[ht]
    \centering
    \includegraphics[width=\linewidth]{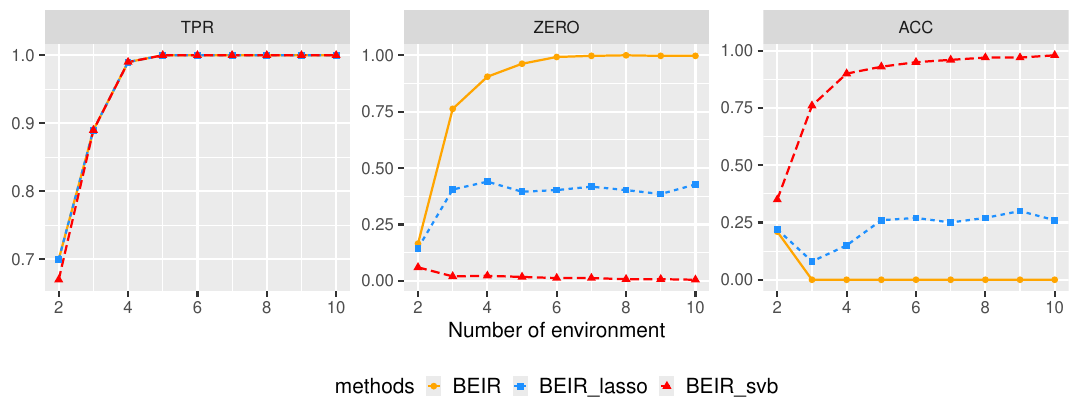}
    \caption{Performance on TPR, ZERO, and ACC from \texttt{BEIR} and two \texttt{BEIR+} variants with $d_Z=4$ irrelevant predictors.}
    \label{fig:set2_zero_beir}
\end{figure}

We compare \texttt{BEIR} with two post-selection variants, \texttt{BEIR\_lasso} and \texttt{BEIR\_svb}, which apply the lasso and spike-and-slab variational Bayes, respectively, as the feature selection method $\mathcal V$ in Algorithm~\ref{alg:beirplus} after the initial \texttt{BEIR} step.
The evaluation metrics in Figure~\ref{fig:set2_zero_beir} include:
the true positive rate for selecting influential invariant predictors (\textbf{TPR}), the fraction of irrelevant predictors incorrectly selected as invariant (\textbf{ZERO}), and the exact recovery of the influential invariant set (\textbf{ACC}).

All \texttt{BEIR}-based procedures achieve TPR close to $1$ as $E$ increases, indicating reliable screening of truly invariant predictors.
However, the naive \texttt{BEIR} retains a non-negligible fraction of irrelevant predictors, reducing exact recovery accuracy.
This behavior is consistent with Section~\ref{sec:irre_predictors}, where the posterior concentrates on an equivalent invariant class that may include elements of $Z$.
Both refinements reduce ZERO and improve ACC.
The \texttt{BEIR\_lasso} improves sparsity but may miss exact recovery in finite samples, due to the lack of general model selection guarantee for lasso \citep{Zhao2006lasso}.
In contrast, \texttt{BEIR\_svb} yields the strongest overall performance and achieves near-exact recovery across a wide range of $E$, owing to the selection consistency of spike-and-slab variational Bayes \citep{Zhang2025Concert}.
We therefore use \texttt{BEIR\_svb}, denoted as \texttt{BEIR+}, in subsequent comparisons.

Table~\ref{tab:setting2_zero_main} reports the comparison with pooled spike-and-slab (\texttt{Pool}), \texttt{EILLS}, and \texttt{BIP} under $d_Z=4$; \texttt{ICP} is removed due to intractability. We further include the false discovery rate among nonzero predictors (\textbf{FDR}) and the $\ell_2$ estimation error of $\bbeta_0$ (\textbf{RMSE}).
Results for $d_Z=8$ are provided in Section~S5.4.

\begin{table}[ht]
\renewcommand{\arraystretch}{1}
\small
\centering
\caption{Comparison under irrelevant predictors with $d_Z=4$. Results are averaged over 100 replications with standard errors in brackets.
Bold values indicate the best performance for each metric, and underlined values indicate the second best.}
\label{tab:setting2_zero_main}
\begin{tabular}{llccccc}
\toprule
$E$ & Method & TPR & FDR & ZERO & ACC & RMSE \\
\midrule
$E=2$ & \texttt{BEIR+} & \textbf{0.670 {\tiny\color{darkgray} (0.422)}} & \textbf{0.130 {\tiny\color{darkgray}(0.284)}} & \textbf{0.060 {\tiny\color{darkgray}(0.118)}} & \textbf{0.350 {\tiny\color{darkgray}(0.479)}} & \textbf{0.417 {\tiny\color{darkgray}(0.269)}} \\
& \texttt{EILLS} & 0.515 {\tiny\color{darkgray}(0.351)} & 0.726 {\tiny\color{darkgray}(0.189)} & 0.158 {\tiny\color{darkgray}(0.194)} & 0.010 {\tiny\color{darkgray}(0.100)} & 0.725 {\tiny\color{darkgray}(0.136)} \\
& \texttt{BIP} & 0.275 {\tiny\color{darkgray}(0.329)} & \underline{0.338 {\tiny\color{darkgray}(0.414)}} & \underline{0.150 {\tiny\color{darkgray}(0.167)}} & \underline{0.030 {\tiny\color{darkgray}(0.171)}} & \underline{0.683 {\tiny\color{darkgray}(0.176)}} \\
& \texttt{Pool} & \underline{0.540 {\tiny\color{darkgray}(0.374)}} & 0.829 {\tiny\color{darkgray}(0.113)} & \underline{0.150 {\tiny\color{darkgray}(0.236)}} & 0.000 {\tiny\color{darkgray}(0.000)} & 0.697 {\tiny\color{darkgray}(0.096)} \\
\addlinespace[0.8ex]
$E=4$ & \texttt{BEIR+} & \textbf{0.990 {\tiny\color{darkgray}(0.070)}} & \textbf{0.000 {\tiny\color{darkgray}(0.000)}} & \underline{0.022 {\tiny\color{darkgray}(0.072)}} & \textbf{0.900 {\tiny\color{darkgray}(0.302)}} & \textbf{0.104 {\tiny\color{darkgray}(0.075)}} \\
& \texttt{EILLS} & 0.220 {\tiny\color{darkgray}(0.328)} & 0.250 {\tiny\color{darkgray}(0.400)} & \textbf{0.003 {\tiny\color{darkgray}(0.025)}} & 0.060 {\tiny\color{darkgray}(0.239)} & 0.658 {\tiny\color{darkgray}(0.181)} \\
& \texttt{BIP} & 0.250 {\tiny\color{darkgray}(0.392)} & \textbf{0.000 {\tiny\color{darkgray}(0.000)}} & 0.040 {\tiny\color{darkgray}(0.099)} & \underline{0.170 {\tiny\color{darkgray}(0.378)}} & \underline{0.568 {\tiny\color{darkgray}(0.240)}} \\
& \texttt{Pool} & \underline{0.530 {\tiny\color{darkgray}(0.354)}} & 0.855 {\tiny\color{darkgray}(0.090)} & 0.045 {\tiny\color{darkgray}(0.097)} & 0.000 {\tiny\color{darkgray}(0.000)} & 0.725 {\tiny\color{darkgray}(0.061)} \\
\addlinespace[0.8ex]
$E=6$ & \texttt{BEIR+} & \textbf{1.000 {\tiny\color{darkgray}(0.000)}} & \textbf{0.000 {\tiny\color{darkgray}(0.000)}} & \underline{0.013 {\tiny\color{darkgray}(0.055)}} & \textbf{0.950 {\tiny\color{darkgray}(0.219)}} & \textbf{0.068 {\tiny\color{darkgray}(0.035)}} \\
& \texttt{EILLS} & 0.145 {\tiny\color{darkgray}(0.304)} & 0.013 {\tiny\color{darkgray}(0.105)} & \textbf{0.003 {\tiny\color{darkgray}(0.025)}} & 0.070 {\tiny\color{darkgray}(0.256)} & 0.635 {\tiny\color{darkgray}(0.170)} \\
& \texttt{BIP} & 0.620 {\tiny\color{darkgray}(0.363)} & \textbf{0.000 {\tiny\color{darkgray}(0.000)}} & 0.025 {\tiny\color{darkgray}(0.075)} & \underline{0.400 {\tiny\color{darkgray}(0.492)}} & \underline{0.360 {\tiny\color{darkgray}(0.257)}} \\
& \texttt{Pool} & \underline{0.995 {\tiny\color{darkgray}(0.050)}} & 0.779 {\tiny\color{darkgray}(0.020)} & 0.185 {\tiny\color{darkgray}(0.224)} & 0.000 {\tiny\color{darkgray}(0.000)} & 0.601 {\tiny\color{darkgray}(0.038)} \\
\bottomrule
\end{tabular}
\end{table}

Overall, \texttt{BEIR+} achieves the best ACC and RMSE across all settings, while maintaining the lowest ZERO and FDR, highlighting its effectiveness in the presence of irrelevant predictors.
Competing methods are less reliable for distinguishing truly invariant predictors and spurious predictors, resulting in higher FDR.

\begin{figure}[ht]
    \centering
    \includegraphics[width=0.8\linewidth]{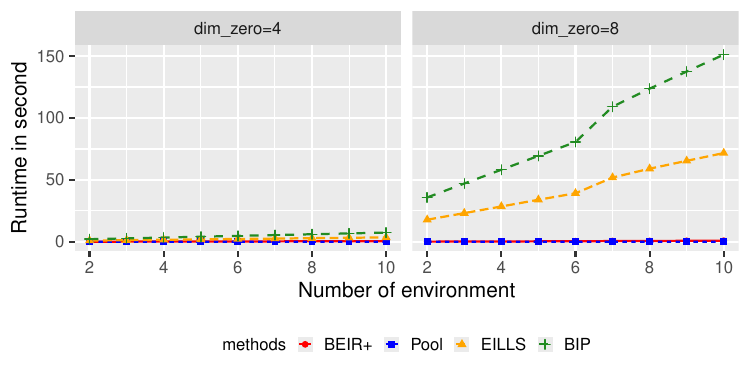}
    \caption{Runtime in second under different dimensions of irrelevant predictors and numbers of environments.}
    \label{fig:setting2_zero_time_main}
\end{figure}

The runtimes under $d_Z=\{4,8\}$ and $E\in\{2,\ldots,10\}$ are provided in Figure~\ref{fig:setting2_zero_time_main}.
Our \texttt{BEIR+} scales favorably and remains comparable to pooling baselines in runtime, while \texttt{EILLS} and \texttt{BIP} become much slower as the number of environments grows and the dimension increases.
This efficiency is mainly due to the simplified working model Eq.\eqref{eq:X_model}, which avoids estimating the dense dependence between $\bx_{I}$ and $\bx_{I^c}$.

\subsection{Analysis of Sachs' protein signaling data}
\label{sec:sachs}

We apply \texttt{BEIR} to the multivariate flow cytometry data of \citet{sachs2005}, which quantify signaling activity in human primary naïve CD4$^+$ T cells under a collection of perturbations.
Each sample consists of measurements on 11 phosphorylated proteins and phospholipids. 
The data comprise 14 perturbations with different stimulatory cues or inhibitory interventions, whose sample sizes range from 707 to 927.

Following standard practice for this benchmark data \citep{Wang2017DAG, Yuan2018DAG}, we treat each perturbation as an environment.
A consensus network based on biological knowledge is available and has been widely used as a reference in causal discovery studies.
We use the verified causal sets implied by this consensus network as the ground truth for evaluating invariant set identification; details are provided in Section~S6.

We focus on ten target proteins that admit environment-invariant causes.
For each target, the remaining ten proteins are used as candidate predictors.
To avoid direct distribution shifts induced by interventions on the response, we exclude environments in which the target is perturbed.

We assess cross-environment generalization using a leave-one-environment-out setup.
In each environmental split, we identify the invariant predictors and fit the corresponding model using the training environments, and then evaluate the predictive performance on the held-out environment.
We compare \texttt{BEIR} with \texttt{EILLS} \citep{Fan2024Environment}, \texttt{ICP} \citep{ICP2016}, and \texttt{BIP} \citep{wu2025bayesian}.
The \texttt{ICP} procedure does not select invariant predictors across significance levels $\alpha\in\{0.001,0.01,0.05,0.1\}$ and is therefore omitted from subsequent comparisons.
Table~\ref{tab:protein_full} reports the target-wise results, excluding \textit{PLCg} and \textit{PIP2} since no invariant set is identified by any method.

\begin{table}[ht]
\renewcommand{\arraystretch}{1}
\setlength{\tabcolsep}{3.8pt}
\footnotesize
\centering
\caption{Invariant predictor identification and leave-one-environment-out prediction for proteins in the Sachs' protein signaling data. Results are averaged over all environmental splits.
In each row, bold values indicate the best performance for each metric, and underlined values indicate the second best.}
\label{tab:protein_full}
\begin{tabular}{lccccccccc}
\toprule
& \multicolumn{3}{c}{\texttt{BEIR}} & \multicolumn{3}{c}{\texttt{BIP}} & \multicolumn{3}{c}{\texttt{EILLS}} \\
\cmidrule(lr){2-4}\cmidrule(lr){5-7}\cmidrule(lr){8-10}
Protein & Precision & Recall & RMSE & Precision & Recall & RMSE & Precision & Recall & RMSE \\
\midrule
\textit{Raf} & \textbf{0.500} & \textbf{0.771} & \textbf{0.824} & \textbf{0.500} & \underline{0.743} & \underline{0.826} & 0.167 & 0.014 & 0.972 \\
\textit{Mek} & \textbf{0.612} & \textbf{0.792} & \textbf{0.759} & \underline{0.595} & \underline{0.778} & \underline{0.763} & 0.250 & 0.014 & 0.804 \\
\textit{Erk} & \underline{0.823} & \textbf{0.755} & 0.972 & \textbf{0.825} & \underline{0.714} & \underline{0.971} & 0.119 & 0.041 & \textbf{0.544} \\
\textit{Akt} & \underline{0.760} & \textbf{0.615} & \textbf{0.816} & 0.731 & \underline{0.531} & \textbf{0.816} & \textbf{1.000} & 0.042 & 0.996 \\
\textit{PKA} & \textbf{0.509} & \textbf{0.808} & \underline{0.932} & \underline{0.480} & \underline{0.731} & 0.934 & 0.000 & 0.000 & \textbf{0.837} \\
\textit{PKC} & \textbf{0.196} & \textbf{0.394} & \underline{1.050} & \underline{0.176} & \underline{0.364} & 1.070 & 0.000 & 0.000 & \textbf{0.948} \\
\textit{p38} & 0.488 & \underline{0.757} & \underline{0.773} & \underline{0.495} & \textbf{0.771} & 0.808 & \textbf{0.500} & 0.529 & \textbf{0.681} \\
\textit{JNK} & \textbf{0.455} & \textbf{0.671} & \underline{0.941} & \underline{0.449} & \underline{0.629} & 0.972 & 0.000 & 0.000 & \textbf{0.939} \\
\bottomrule
\end{tabular}
\end{table}

Overall, \texttt{BEIR} exhibits strong performance across targets.
For most proteins, \texttt{BEIR} achieves the highest precision and recall, indicating effective recovery of biologically supported invariant predictors.
Although \texttt{EILLS} has smaller RMSE for some targets, this often coincides with near-empty selected sets and low precision and recall.
While \texttt{BIP} can be competitive in invariant set identification for some targets, its prediction errors are less consistent across proteins.
These results support the practical benefit of \texttt{BEIR}, which explicitly models environment-specific spurious mechanisms.

\section{Discussion}
\label{sec:discussion}

We propose \texttt{BEIR}, a Bayesian framework for invariant learning from heterogeneous environments, where the joint distribution is explicitly factorized into a shared response mechanism and environment-specific or response-dependent associations.
By forcing predictors to compete between invariant and spurious roles through a competitive spike-and-slab prior, \texttt{BEIR} turns environmental heterogeneity into a source of evidence rather than a nuisance.
At the modeling level, \texttt{BEIR} uses a simplified working model that supports efficient implementation.
At the theoretical level, we show that the working posterior consistently selects the true invariant set and contracts around the true invariant effects. We also show that pooling-based inference can be systematically biased towards selecting spurious predictors.
When irrelevant predictors are present, we establish posterior concentration on an equivalent invariant class and propose a post-selection refinement \texttt{BEIR+} that consistently recovers the minimal invariant set.

Several directions merit further investigation.
First, our current analysis relies on a simplified working specification for computational tractability.
Although the theory and simulations support this choice, richer dependence structures between invariant and spurious predictors could broaden its applicability. 
For example, low-rank factors may capture stronger dependence while preserving tractability.
Second, extension to high-dimensional settings remains important. While \texttt{BEIR+} offers one practical route, it would be valuable to develop unified one-step Bayesian procedures with complexity control, potentially through hierarchical shrinkage priors and scalable variational approximations.
Finally, the framework can be extended beyond Gaussian linear models. Incorporating generalized linear models or nonlinear mechanisms would widen the scope of scientific applications.

\section{Acknowledgments}

This work is supported in part by the fund from the National Nature Science Foundation of China 12331009.

\bibliographystyle{abbrvnat}
\bibliography{refs}
\end{document}